# Thermal Radiation Exchange between Nanoparticles Heated by Arc Discharge

A. Povitsky[1,*] and M.N. Shneider[2]

[1]Department of Mechanical Engineering, The University of Akron, Akron OH USA

[2] Department of Mechanical and Aerospace Engineering, Princeton University, Princeton, NJ USA

[*] alex14@uakron.edu

## Abstract

The heating of particles by plasma radiation plays a critical role in space science involving dusty plasma as well as in industrial processes such as plasma vapor deposition, microchip production, etching and plasma fusion. Numerical modeling of radiation heat transfer from plasma to nano-scale particles includes exchange of scattered thermal radiation between particles-an effect that was neglected in prior studies in which temperature of particles was estimated. Thermal modeling of gas loaded with nanoparticles differs from a typical multiphase flow, where particles are assumed to be in thermal equilibrium with the surrounding gas. In contrast, the temperature of nanoparticles heated by radiation is significantly higher than the local gas temperature. The nanoparticles' volume heating by radiation is markedly different from conventional surface heating experience by macroscale particles. The larger particles are heated to higher temperatures than smaller ones. The study includes numerical modeling of thermal radiation scattered by particles in the Rayleigh regime in where particles' radii are much smaller compared to the radiation wavelength. The study investigates the effects of reduction in conductive heat flux by reducing the gas pressure and using alternating noble gases. Additionally, it investigates the role of enhancement of radiation heat flux from the arc. The computational results show that the re-radiation by larger, heated nanoparticles is important to obtain the accurate temperature of particles. This inter-particle thermal interaction leads to higher temperatures in smaller particles than models assuming thermally isolated particles would predict.

## Introduction

In literature, radiation modeling for dusty gas/plasma in the Rayleigh regime is limited to isolated particles[1,2,3] even though, for higher concentrations of particles, the collective exchange of thermal radiation by multiple scattering is significant[4]. The dependent scattering in an absorbing host medium is still an unsolved problem both theoretically and experimentally. Particularly, plasma particles' sizes — ranging from a nanometer to microns — can significantly affect the propagation of thermal radiation, whose dominant wavelength is approximately 500nm for arc temperature of 7000 K. Most naturally occurring and man-made ensembles of particles are of the range of sizes, and the present study will focus on scattering by particles of different sizes in proximity. The problem under consideration is therefore related to the synthesis of nanoparticles in an arc discharge.

The dusty plasmas are ubiquitous in space[5], including in the vicinity of satellites and space stations. The presence of particles in plasma plays a critical role in industrial processes such as plasma vapor deposition, microchip production, etching and in plasma fusion, to name a few applications. The particles grow in the gas phase through homogeneous chemical reactions and remain trapped in the gas phase. In microelectronic technologies these particles were considered as an important source of irremediable defects[6].

The modeling of gas and plasma loaded with nanoparticles is different from a typical multiphase flow in which particles are in thermal equilibrium with surrounding gas. On the contrary, the temperature of particles heated by radiation from plasma significantly exceeds the local gas temperature[1]. The present study focuses on the radiation scattering of particles in the Rayleigh regime, in which the radius of a particle is much smaller compared to the radiation wavelength. In the considered Rayleigh regime, larger particles are heated to higher temperatures than smaller particles[1-3] — this is because electromagnetic energy absorption is volume-dependent while cooling (dominated by thermal conduction and convection to the ambient gas) is area-dependent. In the present study the absorption of scattered radiation by neighboring particles will be accounted for that assist in accurately predicting the particles' temperatures.

The radiation will be accounted for the grey approximation of radiation intensity. When electromagnetic radiation is illuminated upon a particle, it drives the electric charges in the particle into oscillatory motion that can then radiate secondary electromagnetic waves in all directions. This secondary radiation is called the radiation scattered by the particle. Due to proximity of particles, the scattering rate could vary that will be investigated in the present study. This effect is called dependent scattering: a dependence of the scattering strength on the separation distance between the particles.

Until recently the near field (when the separation between two objects exchanging photons is less than or of the order of the photon wavelength) was inaccessible to experiments, and the interest was focused on far-field effects. Now with advances in nanotechnology and nano-optics, the richness of the near field theory is being exploited. This includes production by scattering high light intensities with spatial variations shorter than the wavelength — a phenomenon that enables rich new physics, both linear and nonlinear, at the nanoscale[7]. Therefore, numerical modeling of radiation heat transfer from gas or plasma to nano-scale particles in Rayleigh regime and exchange of scattered thermal radiation between particles is a promising way to determine their temperature.

For particles exchanging radiation in the Rayleigh regime, the scattering intensity decreases $\sim r^6$ where r is the radius of particles. The radiation absorption intensity is proportional to $r^3$. This implies that, for small particles, homogeneous volume heating by radiation occurs. On the contrary, for larger particles exposed to radiation, only its surface is heated, and the absorbed intensity is proportional to $r^2$. Therefore, in macro-scale engineering radiation problems, emissivity is treated as a surface property. In the Rayleigh regime, thermal absorption and emission are volumetric processes, as the emission emerging from the surfaces really originates from the interior of particles. In the Rayleigh regime, the scattering intensity is ~ $\lambda^{-4}$ while absorption intensity is proportional to $\lambda^{-1}$ , see Refs[8],[9].

Quantifying the exchange of radiation between heated particles in the Rayleigh regime requires the development of an appropriate computational model presented in Section 2. This model includes thermal heating by discharge to isolated particles[1] and re-radiation by nano-scale particles of different sizes. To follow[1-3] the particles are located far from the radiation source (arc discharge) and in a non-ionized gas. The computational results are validated and discussed in Section 3 where the parametric investigation will be conducted to categorize whether the re-radiation by heated particles is important or whether the particles can be assumed to be isolated and interacting with plasma radiation individually. It will be shown that smaller size nanoparticles are heated to a larger temperature by re-radiation by neighboring bigger particles.

## 2. Mathematical model of heating particles

The problem of transient heating balance of an ensemble of particles is considered where the particles are in the arc radiation field, suspended far from its axis, where the buffer gas can be considered non-ionized.

For Rayleigh particles the absorption efficiency is described as follows[10]:

$$c_{abs} = \frac{\pi^2 D^3 E(m)}{\lambda} = \frac{\pi^2 D^3 E(m)\nu}{c}, \tag{1}$$

where $\nu = c/\lambda$ is the radiation frequency; $D$ is particle diameter, $E(m)$ is the function of the complex refractive index $m$, $E(m) = -\mathrm{Im}\left(\frac{m^2-1}{m^2+2}\right)$ We use the broadband value $E(m) \approx 0.35$, which is close to the value estimated[1],[11],[12] for soot particles, $E(m) \sim 0.32 - 0.4$.

Consider the arc radiation as a blackbody radiation with a certain known emissivity $\zeta < 1$ and the Planck blackbody radiation spectral intensity

$$I(\nu) = \frac{2\pi h\nu^3}{c^2}\frac{1}{e^{h\nu/kT}-1}, \tag{2}$$

where $I(\nu)d\nu$ is the radiant power per unit area of radiating surface in the frequency range from $\nu$ to $\nu + d\nu$.

By integrating, the total rate of absorption of the radiation energy by a spherical particle of the diameter $D$ at a distance $r$ from the arc with the temperature $T_{arc}$ is obtained

$$Q_{abs} = \zeta\int_0^\infty c_{abs} I(\nu, r)d\nu = \zeta\,\frac{8}{15}\frac{\pi^7 D^3 E(m) T_{arc}^5}{h^4 c^3}\frac{r_0^2}{r^2} = \zeta\,\frac{4\pi^2 D^3 E(m)\sigma_{SB} k_B T_{arc}^5}{hc}\frac{r_0^2}{r^2}, \tag{3}$$

where $\sigma_{SB} = \frac{2}{15}\frac{\pi^5 k_B^4}{h^3 c^2} \approx 5.670373 \cdot 10^{-8}$ W m$^{-2}$ K$^{-4}$ is the Stefan–Boltzmann constant, $h$=6.62607015 x $10^{-34}$ Js is the Planck constant, c=299792458 m/s is the speed of sound and $k_B$=

1.380649 x $10^{-23}$ J/K is the Boltzmann constant, $r_0$ is the radius of the arc, and r is the distance from the axis of symmetry of the arc to the particle.

For a typical arc discharge used for the synthesis of carbon nanoparticles, the emissivity is $\zeta \approx 0.8$ [1],[13],[14].

Particle cooling occurs mainly in collisions with atoms/molecules of the buffer gas and radiation. The arc radiation will be considered as grey body radiation with temperature $T_{arc}$, a certain known emissivity[1] and the Planck spectral intensity of radiation that heat-up particles in Rayleigh regime with heat flux ~$T^5_{arc}$.

The heated particle in Rayleigh regime is cooled by radiation $Q_{rad}$ to gas ambiance with the radiative heat losses ~$T_p^5$ (see Ref[1] and references therein), determined by the expression[15],[16]

$$Q_{rad} = 4\pi a^2 \int_0^\infty \varepsilon_\lambda \frac{2\pi h c^2 d\lambda'}{\lambda'^5\left[exp(hc/\lambda' k_B T_p)-1\right]}, \tag{4}$$

where $\varepsilon_\lambda = 8\pi a E(m)/\lambda$ is the emissivity of the particle and $a=D/2$ is particle radius. Integrating the right-hand side of Eq. (4) over the entire frequency spectrum, we find a formula for the power of the radiation energy losses:

$$Q_{rad} = \frac{16\pi^2 D^3 E(m)\sigma_{SB} k_B T_p^5}{hc}. \tag{5}$$

Along with the radiation cooling, a particle is heated by the thermal radiation from the background gas with the temperature $T_g$. However, the nobble gas emissivity $\zeta_g << 1$ and the heating by the surrounding gas could be neglected.

The particle is also cooled by conduction in the collisions with the buffer gas atoms. The rate of heat loss in these collisions[1,17] is $Qg$ and depend on the temperatures of particle, $T_p$, and surrounding gas, $T_g$.

$$Q_g = \frac{2\pi a^2 \alpha_T p_g}{T_g}\sqrt{\frac{R_m T_g}{2\pi\mu}}\left(\frac{\gamma+1}{\gamma-1}\right)(T_p - T_g). \tag{6}$$

For helium as a buffer gas[1] used in this study $\mu_{He} = 4$ g/mole; $\gamma = 1.66$; and $R_m = 8.314$ J/mole K. Hear $\alpha_T$ is the thermal accommodation coefficient of ambient gases with the surface of a particle. The exact value of the accommodation coefficient can be determined only by comparison of the theoretical calculations with the results of measurements. In this paper, we take $\alpha_T \approx 0.1$, given in[1],[18],[19] for the carbon nanoparticles in helium.

In the non-ionized gas, at a distance away from the arc discharge the particle loses electrons through the thermionic emission, acquiring a positive charge. In this situation, the electron current from the plasma on the particle is negligible. The thermionic emission results in an additional cooling of the nanoparticles, $Q_{TE}$, because each "evaporating" electron carries away the energy with the heat loss formula the thermionic emission current presented in[1],[20].

For the non-ionized gas or weakly ionized plasma, at the distances of centimeters from the arc discharge when the electron current from the plasma on the particle is negligible as compared to the current thermionic emission, the particle loses electrons through the thermionic emission, acquiring a positive charge

$$q = \int_t I_{e,T}\, dt, \tag{7}$$

and the potential

$$\phi = q/C > 0, \tag{8}$$

where $C = 4\pi\varepsilon_0 a$ is the capacity of the particle and $\varepsilon_0$= 8.85×10$^{-12}$ F·m$^{-1}$ is vacuum permittivity..

In this case, the positive potential suppresses the thermionic current and is equal to:

$$I_{e,T} = 4\pi a^2 A T_p^2\, exp[-(w_a + \phi)/kT_p]. \tag{9}$$

The thermionic emission results in an additional cooling of the nanoparticles, because each "evaporating" electron carries away the energy $(w_a + \varphi)$. The corresponding power of heat loss measured in eV/s is

$$Q_{TE} = \left(\frac{|I_{e,T}|}{e}\right)(w_a + \varphi + 2k_B T_p) \tag{10}$$

where $e$ =1.602x10$^{-19}$ Coulomb is the elementary charge, defined as the electric charge carried by a single electron, $A = 120$ A/cm$^2$K$^2$, $w_a = 4.7$ eV is the work function for the carbon nanoparticle

The energy balance of an individual particle is determined by the following equation[1]:

$$M_p\, C_p \frac{dTp}{dt} = Q_{abs} - Q_{TE} - Q_g - Q_{rad}. \tag{11}$$

Here $M_p = \frac{4}{3}\pi\rho_p a^3$ is the mass of the particle.

The heat balance equation (11), together with (3), (5), (6), (10), which determine heat fluxes in above equation, is solved for the initial conditions at t=0: $q = 0$, $\varphi = 0$; $T_p$=$T_g$. The equation (13) is solved numerically using the explicit Euler method for each of the considered particles of different sizes as they have different temperature, $Tp$. The radiation exchange between particles, which was neglected in prior studies[1-3] is accounted for through the radiation $Q_{rad}$ from particle to gas ambiance multiplied by view factor F between particles (see Eq. (A-2) from Appendix).To obtain $Q_{rad}$ emitted to a present particle by multiple particles, N, of the same size located at the same distance from a present particle, the $Q_{rad}$ in multiplied by N.

## 3. Results and Discussion

The objective is to compute radiation thermal fluxes between particles and dynamics of temperature of particles in Rayleigh regime, $r/\lambda \leq 0.1$, where r is the particle radius and $\lambda$ is wavelength. The goal is to evaluate the importance of re-radiation for evaluation of particles' temperature. To reiterate, the arc temperature $T_{arc}$=7000K, radius of the arc is 5 mm while the distance between the axis of symmetry of the arc and particle is 15 mm, at which distance the irradiated particles are in a non-ionized gas rather than plasma.

(as in Ref[1]). Density of particle material is 2660 kg/m$^3$ and its heat capacity is 1900 J/(kg K) [1,21,22]. The calculations were performed for pressure $p = 68$ kPa[1,23].

In Fig.1, the obtained temperature for particles of 5nm and 50nm radius (their diameters are 10nm and 100nm respectively) is compared to those obtained in Ref[1] for validation purposes. In discussion about Fig. 1 and consequent figures, these particles are denoted as "small" and "big" particles, respectively. To follow Ref[1], the particles are submerged in gas at 1500K. They are heated by plazma arc discharge by radiation at 7000 K and cooled by gas through heat conduction as described mathematically in previous Section. The big particle (blue curve in Fig. 1) is heated to a larger temperature compared to a small particle.

The temperature profiles in Fig. 1 coincide with those shown in Fig. 2[1]. The equilibrium (stationary) temperatures at T=0.0001 sec are 1877.16K for a bigger particle and 1539.31K for a smaller particle. By[1], the stationary temperatures are similar: 1539.47 K for 5 nm particles and 1878.87 K for 50 nm particles.

The simulated time interval T is split to 10000 steps of equal size so as the time step for numerical integration of Eq. (13) is $10^{-8}$ sec. To verify convergence of numerical procedure with the time step, the time step was increased 10 times. The temperature profiles obtained do coincide and equilibrium temperature remains practically the same: 1877.13 and 1539.31 for bigger and smaller particles, respectively.

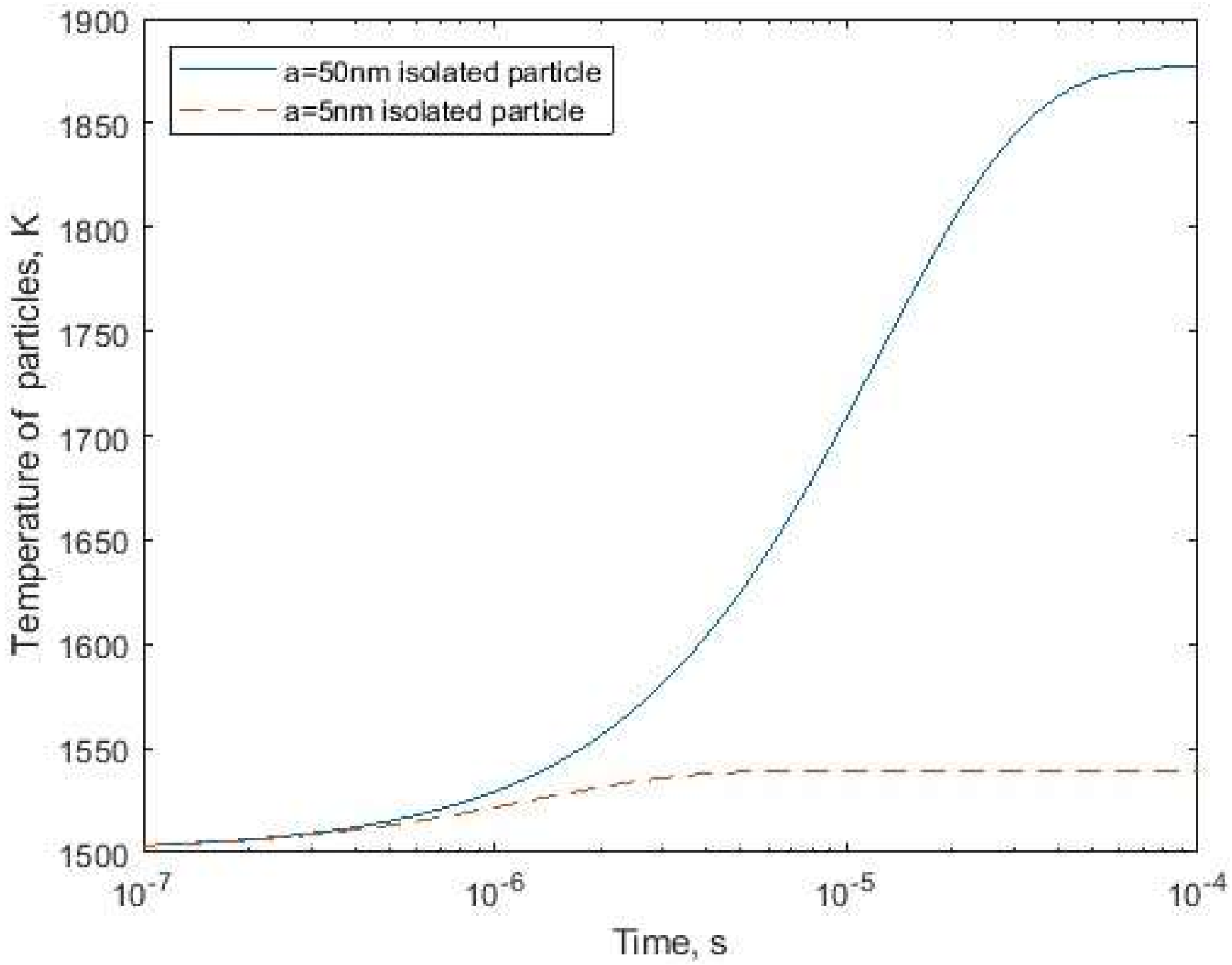

Figure 1: Particles of radius 5 and 50 nanometers submerged in gas at 2000K and exposed to arc radiation at 7000K. Radiation heat exchange between particles is not accounted for.

Next, the role of re-radiation between particles is evaluated when the small particle is located at the center of cloud of particles. Recall that if particles are isolated or located far from each other, they do not exchange radiation between them. At inert gas temperature of 2000K, the big and small particles are heated to equilibrium temperatures of ~2371K and 2042K , respectively (see continous blue and dashed red curves in Fig 2*)*. The transition periods from initial temperature to equilibrium steady temperature for big and small particles are $\sim 8x10^{-5}$ and $\sim 6x10^{-6}$ sec, respectively. It is assumed that equilibrium temperature is established at time moment t when the temperature becomes within 1K from its equibrium value.

In computational results accounting for radiation between particles presented as dashed-dotted curve in Fig. 2, the individual big and small particles are located at distance of 500 nm equal to ten times radius of a bigger particle. This geometric configuration gives the view factor from the smaller to larger particle used in computations, see Ref[24] and Appendix.The temperature of the big particle does not change compared to prior case as the radiation flux from small particle is negligibly small for big particle. Note that the wavelength corresponding to the dominant radiation at temperature $T_{rad}$ is by Wien's law, $\lambda_{dom} \sim 1/T_{rad}$. The dominant wavelength range at 7000 K corresponds to ~500 nm, that is the separation distance between big and small particles is comparable to radiation wavelength.

The temperature of small particle increased only by 0.2 K compared to prior case of neglected radiation between particles that makes the re-radiation effect small for two particles at relatively big distance.

To evaluate the role of re-radiation between cloud of particles and an individual particle, the small particle is located at the center of volume (cloud of particles) of its size $(2\mu)^3$. The number density of big particles is assumed 1.25 x $10^{19}$ that corresponds to 100 particles located within the cloud. To simplify computational model, these 100 big particles are located at equal distance from the small particle equal to half distance between the small particle and edge of the cubical volume, that is, the big particles are located at sphere surface with its radius of 500 nm, where the small particle is located at the center of the sphere.

Dotted curve in Fig. 2 present the case of small particle radiated by 100 big particles located at distance of 500 nm. The temperature of big particles remains the same whereas the small particle is heated up by moderate ~23K and reach ~2065.16K . The transition time for a small particle has increased to ~5x$10^{-5}$ sec compared to heating of isolated particle.

When the separation distance between two objects is less than the characteristic thermal wavelength, the near-field radiative heat flux can exceed the Planckian black-body limit heat flux due to the contribution of evanescent wave tunneling (see Ref[25] and references therein). This super-Planckian effect includes the peculiarities of how particles' radiation influences each other in the near zone and will be considered in future research.

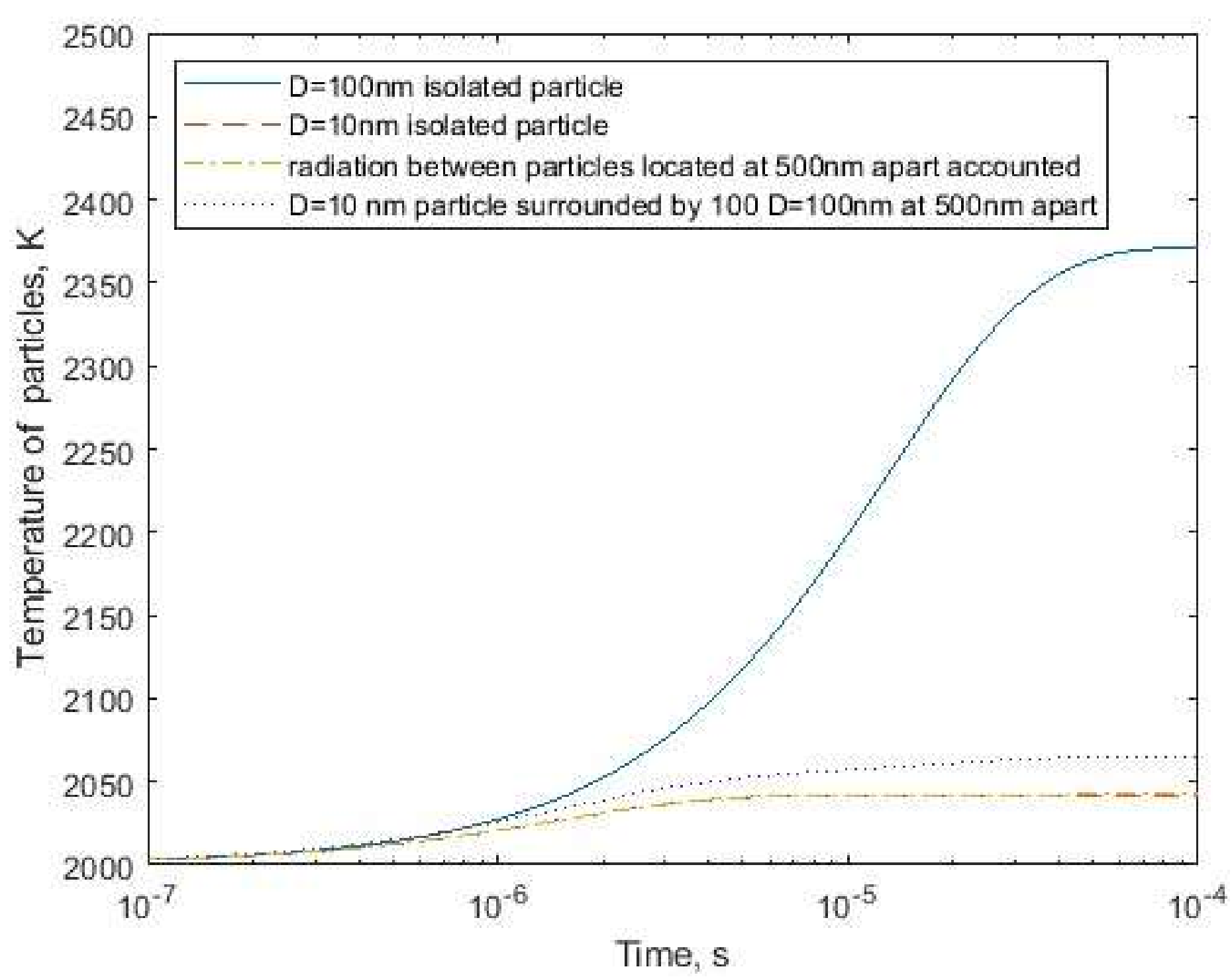

Figure 2:Particles of diameters 10 and 100 nanometers submerged in gas at 2000K and exposed to arc radiation at 7000K. The following situations are presented: (i) no radiation heat exchange between particles; (ii) individual big and small particles with radiation exchange; and (iii) a small particle surrounded with 100 big particles located at the distance of 500 nm with radiation exchange.

To explore conditions in which the effect of re-radiation is more prominent, two situations are modeled and depicted in Fig. 3.The number density is doubled in both cases: (i) the size of cloud is halved to 1μ (dotted curve in Fig.3) and (ii) the number of particles is doubled to 200 particles (dashed-dotted curve in Fig. 3). The equilibrium temperature of a small particle does increase to 2135.14 K and to 2088.27K for cases (i) and (ii), respectively. This corresponds to the increase of equilibrium temperature compared to an isolated particle by about 100K and 50K for cases (i) and (ii), respectively.

It should be noted that case (i) corresponds to distance of 250 nm between small and big particles that is less than the characteristic thermal wavelength. Therefore, the near-field radiative heat flux can exceed the Planckian heat flux[25]. Presented computations for this case correspond to an estimate of a low limit of particle temperature.

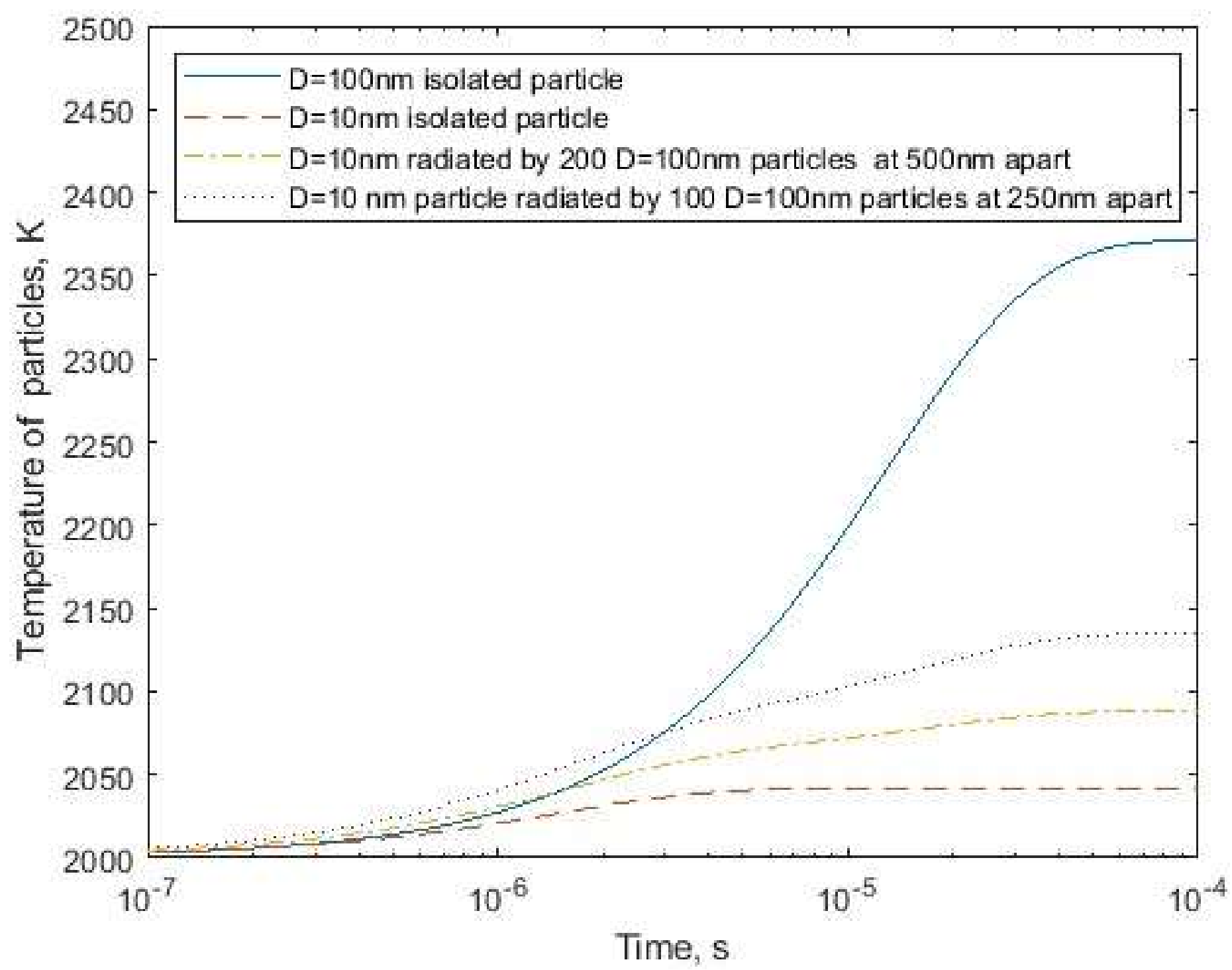

Figure 3: Number density is doubled compared to prior figure. The following situations are presented: the size of particles' cloud is halved, and the number of particles is doubled to 200 particles. Temperature of isolated particles is presented for comparison.

In Figure 4 computations were conducted for reduced gas pressure (34kPa and 17kPa). As in Figure 2, particles of diameters 10 and 100 nanometers are submerged in gas at gas temperature of 2000K and exposed to arc radiation at 7000K. The conductive heat flux between gas and particles is proportional to gas pressure per Eq. (6). For halved pressure p=34kPa the isolated big and small particles are heated to equilibrium temperatures of ~2616K and 2083K , respectively (see dashed-dotted and continous blue curves in Fig 4*).* If radiation between particles is accounted for, the equilibrium temperature of small particle increases to 2162K. If the pressure is halved again to 17kPa, equilibrium temperatures of ~2860K and 2162K are reached for isoltaed big and small particles, respectively (red curves in Fig. 4). If radiation between particles is accounted for, the equilibrium temperature of small particle increases to 2380K, that is, the temperature of small particle is increased by more than 200 K. To sum-up, for lower gas pressures the effect of thermal radiation between particles becomes stronger than that depicted in Figure 2.

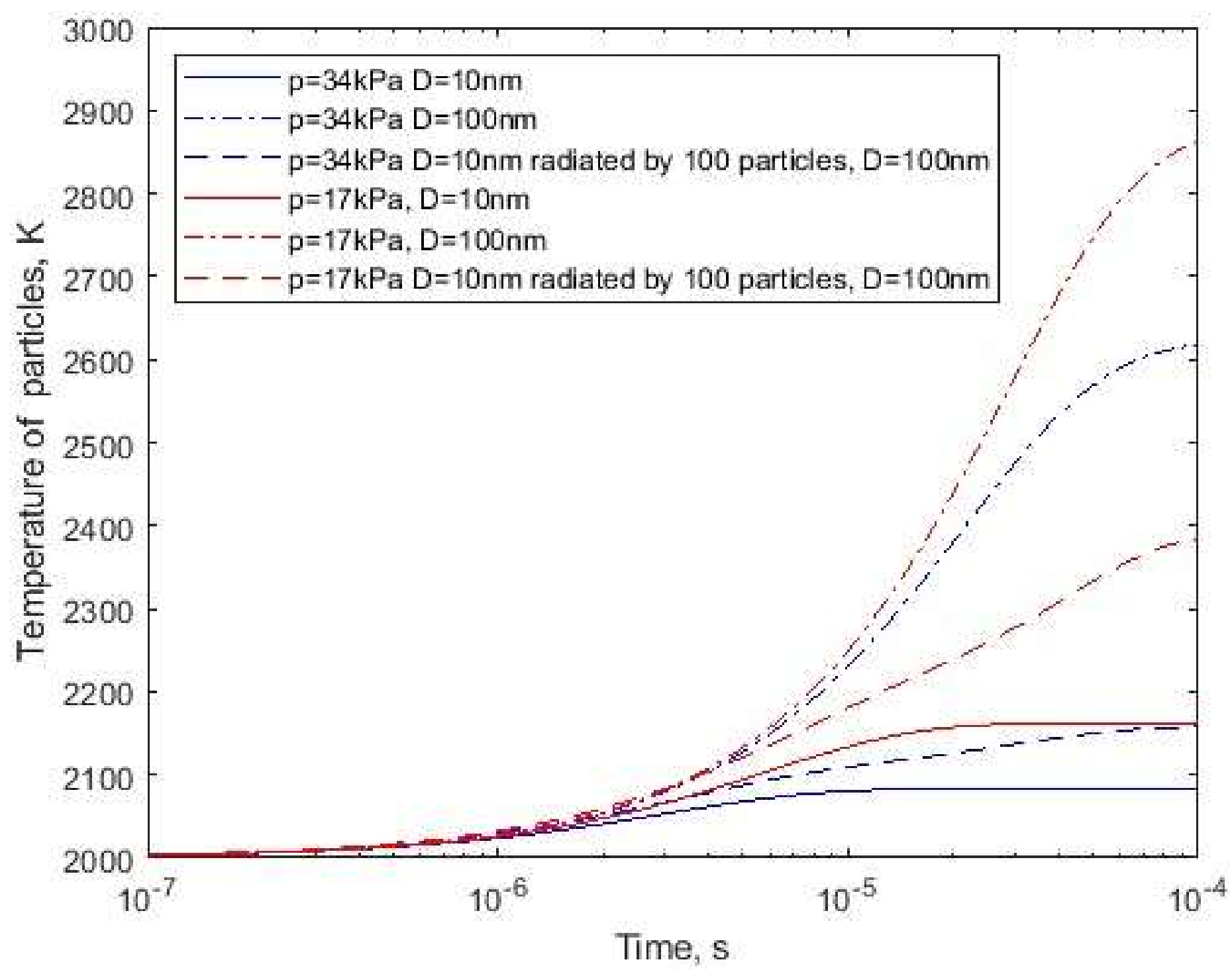

Figure 4: Effects of reduced gas pressure on the importance of thermal radiation between particles

Next, a different way to reduce the magnitude of conductive heat flux by increasing molecular weight of gas is explored. Two noble gases, argon (molecular weight $\mu = 40$ g) and xenon (molecular weight of 131 g) were used in place of helium. By Eq. (6) conductive heat flux is proportional to $1/\sqrt{\mu}$ and, therefore, conductive heat flux decreases with the increase of molecular weight of gas. As in Figure 2, particles of diameters 10 and 100 nanometers are submerged in gas with pressure of 68 kPa and temperature of 2000K and exposed to arc radiation at 7000K. For argon gas, the isolated big and small particles are heated to equilibrium temperatures of ~2785K and 2130K , respectively (see continous and dashed-dotted blue curves in Fig 5*)*.If radiation between particles is accounted for, the temperature of small particles reach 2286K (dashed-dotted blue curve in Fig. 5), that is, the small particle is heated by additional temperature of more than 150K if radiation between particles is accounted for. For xenon, the temperature of small particles are 2226 K and 2577 K, without and with account for radiation between particles, respectively. Account for radiation between particles does increase the temperature of small particles by larger amount of ~350K for xenon gas.

To summarize, the reduction in magnitude of conductive heat flux between particles and surrounding gas increases the importance of accounting for radiation between particles.

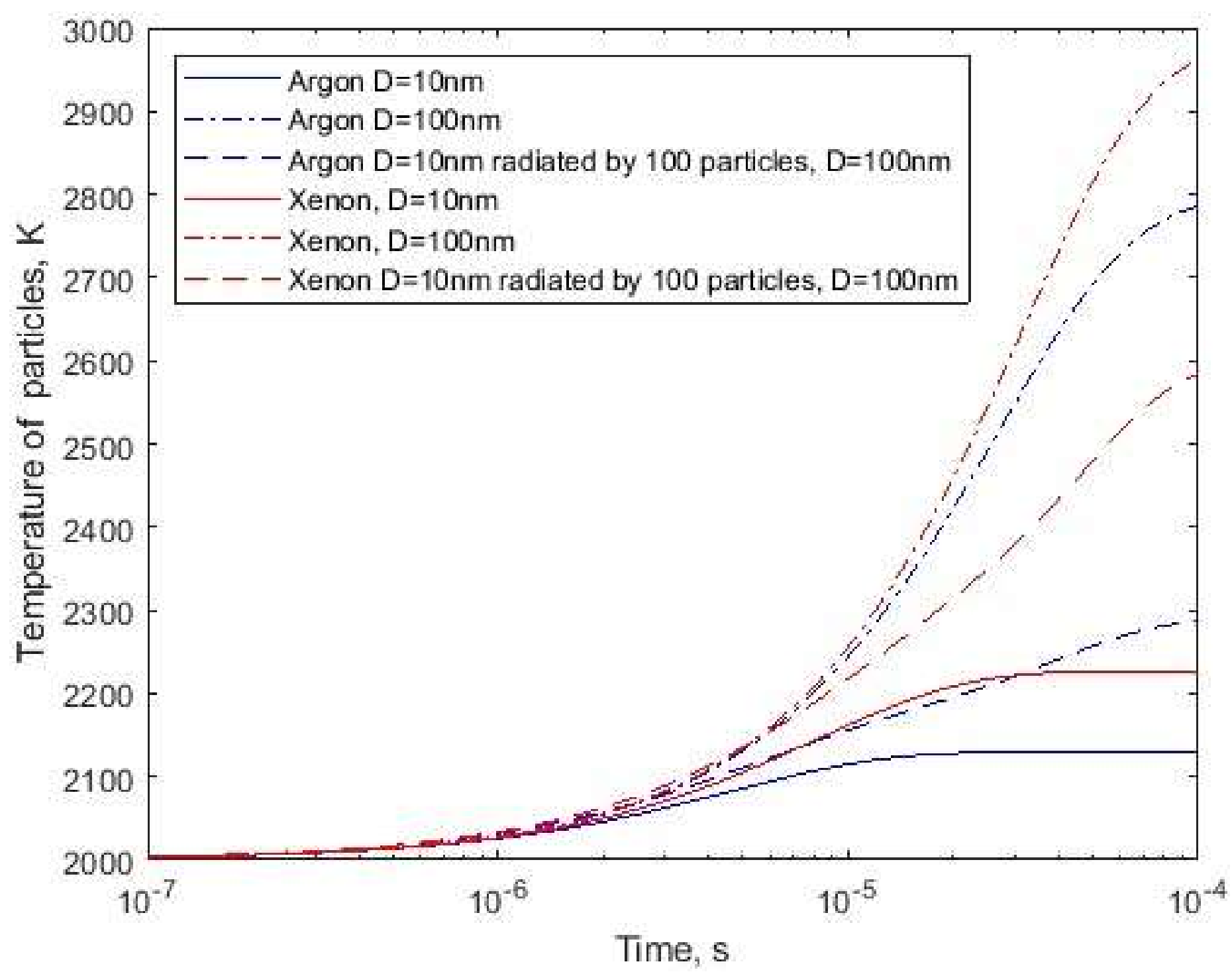

Figure 5: Effects of molecular weight of noble gas on the importance of thermal radiation between particles

To explore the role of effect of arc radiation temperature on radiation heat flux between particles, the arc temperature is increased from 7000K to 8000K and to 9000K. By equation (3) the radiation heat flux from arc is proportional to $T_{arc}^5$, therefore, the increase of temperature to 8000K leads to an increase of heat flux from arc by 1.94 times while the increase of temperature to 8000K leads to an increase of heat flux from arc by 3.51 times. As in Figure 2, particles of diameters 10 and 100 nanometers are submerged in helium gas with its pressure of 68 kPa and temperature of 2000K. For the arc temperature of 8000K, the temperature of small particles increases from 2085K for isolated particles to 2132K when radiation between particles is accounted for. This increase of 47K is doubled compared to the arc temperature of 7000K depicted in Fig. 2. For larger arc temperature of 9000K, the temperature of small particles increases from 2157K to 2262K when radiation between particles is accounted for. This represents the increase of temperature of small particles by more than 100K when radiation between particles is accounted for.

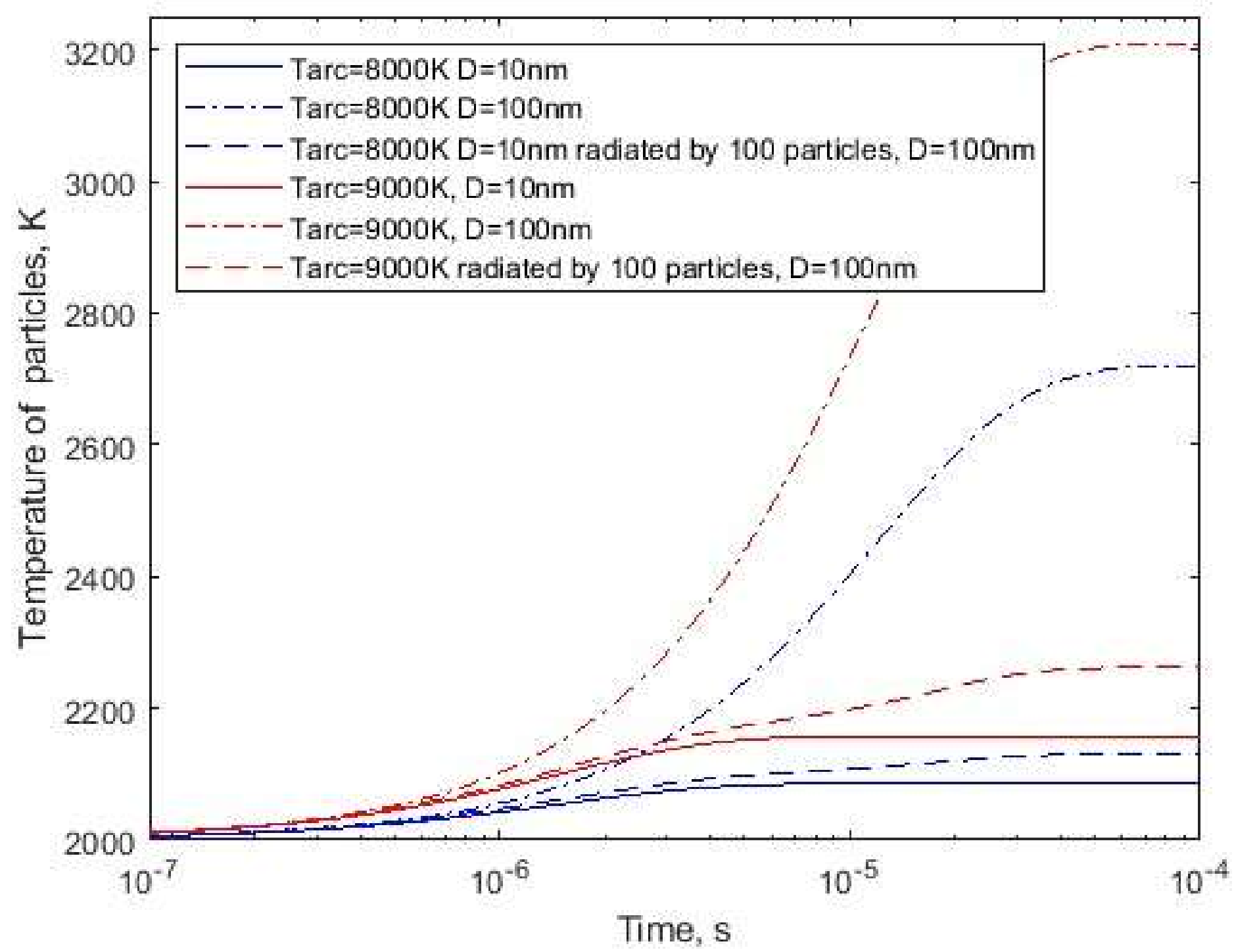

Figure 6: Effect of arc radiation temperature on the importance of thermal radiation between particles

**Conclusions and Future Research**

The model of radiation exchange between nano-scale particles of different sizes was created and validated. The particles are heated by arc discharge and located in non-ionized gas at some distance from the arc. The particles exchange heat with surrounding gas by conduction. The particles absorb and emit thermal radiation by their volumes as opposed to macro-scale particles that exchange radiation by their surfaces. As a result, the bigger nanoparticles are heated to a larger temperature compared to smaller nanoparticles.

The prior studies considered particles as isolated and not exchanging thermal energy by radiation. The present study shows that account for thermal radiation between particles does increase the temperature of smaller particles that absorb radiation emitted by bigger particles. The strength of the effect depends on the density of particles suspended in gas.

The effect of additional heating by radiation between particles is softened by the presence of heat conduction between gas and particles. It is shown that reduction in gas pressure, which reduces the magnitude of conductive heat losses by particles, does increase the significance of the effect. The increase of molecular weight of gas by changing from helium to heavier noble gases, argon and xenon, also makes the effect of heating of smaller particles stronger.

The effect of radiation between particles becomes stronger if the arc temperature increases from 7000K to 8000K and 9000K.

In future research the obtained results in terms of particles' temperature could be used to determine the change in radiation flux through the cloud of particles, gas heating by particles and phase change (melting) of particles leading to particles' sticking to each other. If the particles' number density is high enough or the layer is thick enough, it can block some or all the radiation from the source.

The research could be extended to distances between particles that are smaller than radiation wavelength to account for intense non-Planckian radiation scattering in the near field.

## Appendix: Computations of view factor

The view factor (see Fig. A-1) [24] is the portion of the radiative heat flux, which leaves spherical particle $A_1$, that strikes spherical particle $A_2$. The view factor measures how well one particle can see another particle. View factors are purely geometrical parameters and are independent of the physical surface properties and temperature and absorption of radiation by ambiant gas.

$$F_{d1-2} = 0.5(1 - \sqrt{1 - R^2}), \qquad \text{(A-1)}$$

where R=a/h, a is the radius of bigger particle and h is the distance between particles

By reciprocity rule, the view factor from $A_2$ to $dA_1$ is

$$F_{2-d1} = (a_1/a_2)^2 F_{d1-2} \qquad \text{(A-2)}$$

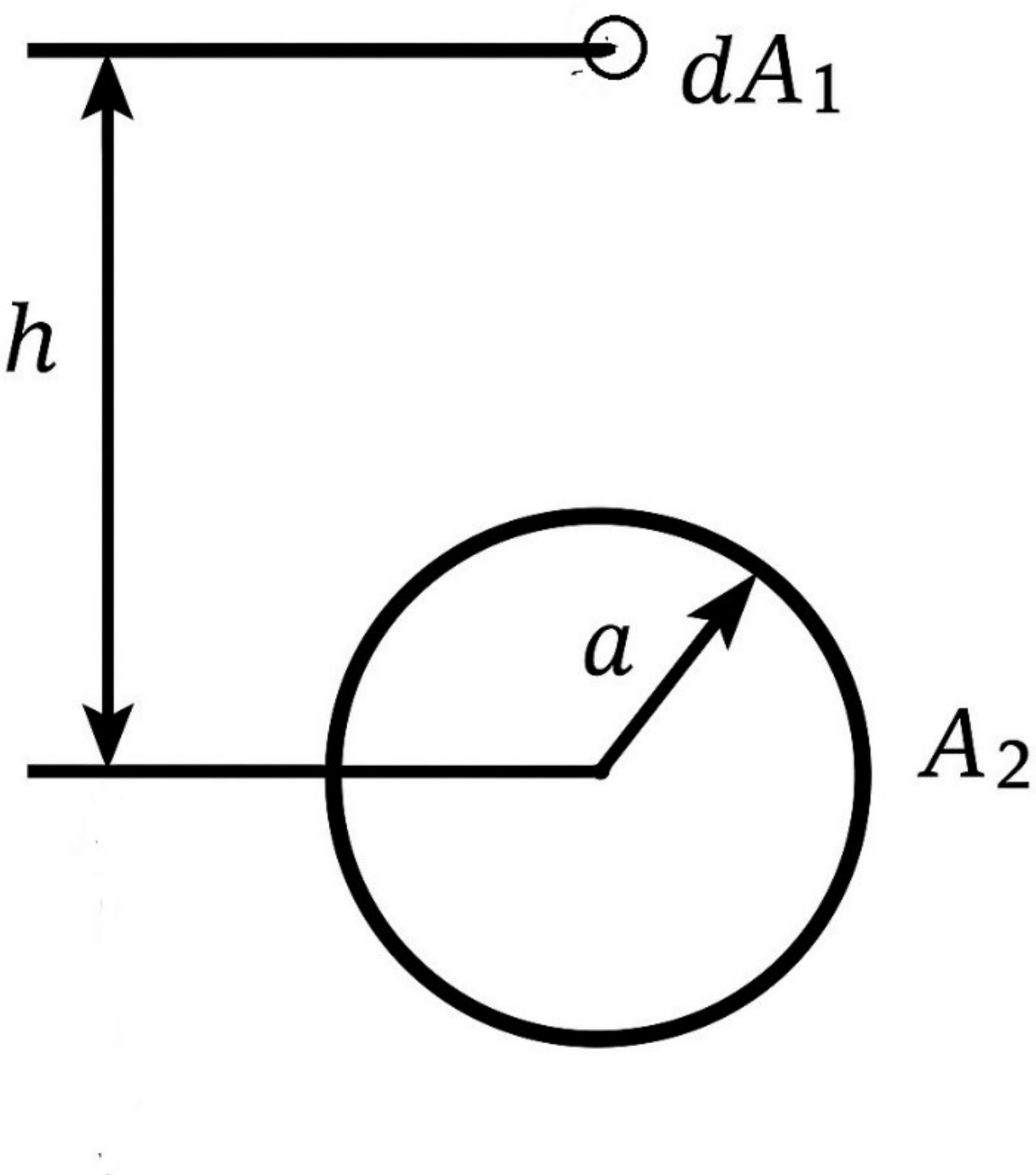

Figure A-1: Radiation view factor from small sphere $dA_1$ to large sphere $A_2$.

**Acknowledgements**

This work of AP was supported by the Department of Energy, Office of Science (Grant No. DE-SC0025441). MS acknowledges support from the Princeton Collaborative Low Temperature Plasma Research Facility (PCRF).

**References**

---

[1]M. N. Shneider, Carbon nanoparticles in the radiation field of the stationary arc discharge, Physics of Plasmas 22, 073303 (2015).
[2] J. M. Mitrani and M. N. Shneider, Time-resolved laser-induced incandescence from multiwalled carbon nanotubes in air, Applied Physics Letters 106, 043102 (2015).

[3] V. Vekselman, Y. Raitses, and M. N. Shneider, Nanoparticles growth in dynamic plasma, Phys. Rev. E, 99, 063205 (2019).
[4] B. X. Wang and C. Y. Zhao, The dependent scattering effect on radiative properties of micro/nanoscale discrete disordered media, Annual Review of Heat Transfer, October 2020. Preprint arXiv:2003.13288v1 [cond-mat.mes-hall]. 30 Mar 2020.

[5] V.E. Fortov et al, Complex (dusty) plasmas: Current status, open issues, perspectives, Physics Reports 421, pp. 1 – 103 (2005).
[6] L Boufendi et al, Dusty plasma for nanotechnology, Journal of Physics D: Applied Physics 44(17):174035 (2011).
[7] X. Fan, W. Zheng and D. J. Singh, Light scattering and surface plasmons on small spherical particles, Light: Science & Applications (2014) 3, e179.
[8] J. H. Seinfeld, and S. N. Pandis, Atmospheric Chemistry and Physics, Chapter 15.1.1; 2nd Edition, John Wiley and Sons, New Jersey, 2006, ISBN 0471720186.
[9] A.J. Cox, An experiment to measure Mie and Rayleigh total scattering cross sections, American Journal of Physics. 70 (6): 620 (2002).

[10] W.W. Stoffels et al, Detection of dust particles in the plasma by laser-induced heating, J. Vac Sci Technology, A 14(2) (1996).
[11] H. A. Michelsen, Understanding and predicting the temporal response of laser-induced incandescence from carbonaceous particles, J. Chem. Phys. **118**, 7012 (2003)

[12] D.R. Snelling, F. Liu, G.J. Smallwood, Ö.L. Gülder, Determination of the soot absorption function and thermal accommodation coefficient using low-fluence LII in a laminar coflow ethylene diffusion flame. Combustion and Flame 136 180–190 (2004).

[13] J. Ng and Y. Raitses, Self-organization processes in the carbon arc for nano synthesis, J. Appl. Phys. 117, 063303 (2015).

[14] A. G. Ostrogorsky and C. Marin, Heat transfer during production of carbon nanotubes by the electric-arc process, Heat Mass Transfer 42, 470–477 (2006).

[15] H. A. Michelsen, Understanding and predicting the temporal response of laser-induced incandescence from carbonaceous particles, J. Chem. Phys. **118**, 7012 (2003)

[16] H.A. Michelsen, F. Liu, B.F. Kock, H. Bladh, A. Boiarciuc, M. Charwath, T. Dreier, R. Hadef, M. Hofmann, J. Reimann, S. Will, P.-E. Bengtsson, H. Bockhorn, F. Foucher, K.-P. Geigle, C. Mounaım-Rousselle, C. Schulz, R. Stirn, B. Tribalet, R. Suntz, Modeling laser-induced incandescence of soot: a summary and comparison of LII models, Appl. Phys. B **87**, 503 (2007)

[17] C. F. Bohren and D. R. Huffman, Absorption and Scattering of Light by Small Particles ,Wiley, New York, 1983.
[18] K. J. Daun, G. J. Smallwood, F. Liu, Investigation of Thermal Accommodation Coefficients in Time-Resolved Laser-Induced Incandescence, J. of Heat Transfer, 130, 121201-32008 (2008).

[19] T. A. Sipkens, R. Mansmann, K. J. Daun, N. Petermann, J.T. Titantah, M. Karttunen, H. Wiggers, T. Dreier, C. Schulz, In situ nanoparticle size measurements of gas-borne silicon nanoparticles by time-resolved laser-induced incandescence, Appl. Phys. B 116, 623 (2014).
[20] J. Orloff, Schottky emission. *Handbook of Charged Particle Optics* (2nd ed.) CRC Press (2008)

[21] L.A. Melton, Soot Diagnostics Based on Laser Heating, Appl. Opt. 23, 2201 (1984).
[22] H.A. Michelsen, F. Liu, B.F. Kock, H. Bladh, A. Boiarciuc, M. Charwath, T. Dreier, R. Hadef, M. Hofmann, J. Reimann, S. Will, P.-E. Bengtsson, H. Bockhorn, F. Foucher, K.-P. Geigle, C. Mounaım-Rousselle, C. Schulz, R. Stirn, B. Tribalet, R. Suntz, Modeling laser-induced incandescence of soot: a summary and comparison of LII models, Appl. Phys. B **87**, 503 (2007)

[23] M. Kundrapu, M. Keidar, Numerical simulation of carbon arc discharge for nanoparticle synthesis, Phys. Plasmas 19, 073510 (2012)

[24]R. Siegel and J. Howell, Thermal Radiation Heat Transfer, 4th Edition, Taylor and Francis (2002).

[25] B. Liu, M. Luo, J. Zhao, L. Liu, and M. Antezza, Many-body interaction on near-field radiative heat transfer between two nanoparticles caused by proximate particle ensembles, International Journal of Heat and Mass Transfer, Vol. 215, 124462 (2023).